\documentclass[aps,prl,preprint,superscriptaddress,showpacs,showkeys]{revtex4-1}

\usepackage{graphicx}
\usepackage{dcolumn}
\usepackage{bm}
\usepackage{color}

\begin{document}


\title[Sample title]{Nanosecond spin-transfer over tens of microns in a bare GaAs/AlGaAs layer}

\author{L. N\'{a}dvorn\'{i}k}
 \email{nadvl@fzu.cz}
 \affiliation{Institute of Physics ASCR, v.v.i., Cukrovarnick\'{a} 10, 16253 Praha 6, Czech Republic}
 \affiliation{Faculty of Mathematics and Physics, Charles University, Ke Karlovu 3, 12116 Praha 2, Czech Republic}
\author{P. N\v{e}mec}
 \affiliation{Faculty of Mathematics and Physics, Charles University, Ke Karlovu 3, 12116 Praha 2, Czech Republic}
 \author{T. Janda}
 \affiliation{Faculty of Mathematics and Physics, Charles University, Ke Karlovu 3, 12116 Praha 2, Czech Republic}
 \affiliation{Institute of Physics ASCR, v.v.i., Cukrovarnick\'{a} 10, 16253 Praha 6, Czech Republic}

\author{K. Olejn\'{i}k}
 \affiliation{Institute of Physics ASCR, v.v.i., Cukrovarnick\'{a} 10, 16253 Praha 6, Czech Republic}
\author{V. Nov\'{a}k}
 \affiliation{Institute of Physics ASCR, v.v.i., Cukrovarnick\'{a} 10, 16253 Praha 6, Czech Republic}
 \author{V.~Skoromets}
 \affiliation{Institute of Physics ASCR, v.v.i., Na Slovance 2, 18221 Praha 8, Czech Republic}
  \author{H.~N\v{e}mec}
 \affiliation{Institute of Physics ASCR, v.v.i., Na Slovance 2, 18221 Praha 8, Czech Republic}
  \author{P. Ku\v{z}el}
 \affiliation{Institute of Physics ASCR, v.v.i., Na Slovance 2, 18221 Praha 8, Czech Republic}
 \author{F. Troj\'{a}nek}
 \affiliation{Faculty of Mathematics and Physics, Charles University, Ke Karlovu 3, 12116 Praha 2, Czech Republic}
\author{T. Jungwirth}
 \affiliation{Institute of Physics ASCR, v.v.i., Cukrovarnick\'{a} 10, 16253 Praha 6, Czech Republic}
 \affiliation{School of Physics and Astronomy, University of Nottingham, Nottingham NG7 2RD, UK}
 \author{J. Wunderlich}
 \affiliation{Institute of Physics ASCR, v.v.i., Cukrovarnick\'{a} 10, 16253 Praha 6, Czech Republic}
 \affiliation{Hitachi Cambridge Laboratory, J. J. Thomson Avenue, CB3 0HE Cambridge, UK}
 

\date{\today}

\begin{abstract}
The spin-conserving length-scale is a key parameter determining functionalities of a broad range of spintronic devices including magnetic multilayer spin-valves in the commercialized magnetic memories or lateral spin-transistors in experimental spin-logic elements. Spatially resolved optical pump-and-probe experiments in the lateral devices allow for the direct measurement of the lengthscale and the time-scale at which spin-information is transferred from the injector to the detector. Using this technique, we demonstrate that in an undoped GaAs/AlGaAs layer spins are detected at distances reaching more than ten microns from the injection point at times as short as nanoseconds after the pump-pulse. The observed unique combination of the long-range and high-rate electronic spin-transport requires simultaneous suppression of mechanisms limiting the spin life-time and mobility of carriers. Unlike earlier attempts focusing on elaborate doping, gating, or heterostructures we demonstrate that the bare GaAs/AlGaAs layer intrinsically provides superior spin-transport characteristics whether deposited directly on the substrate or embedded in complex heterostructures. 
\end{abstract}

\pacs{72.25.Dc, 72.25.Fe}
\keywords{spin transport in semiconductors, time and spatially-resolved spin detection}
\maketitle



In modern magnetic memory bits, spin currents are driven through vertical stacks from  a magnetic polarizer via a non-magnetic spacer into a recording magnet to facilitate  reading and writing of the information \cite{Chappert2007}. Spin conservation across the spacer is an essential prerequisite for these functionalities. Recent discoveries showed that a non-magnetic layer alone can be turned into an efficient injector of the spin-current to the adjacent magnetic layer by relativistic electrical transport means, namely by the spin Hall effect \cite{Sinova2014}. Here the spin-conserving length-scale defines the active width of the non-magnetic layer from which the spin-current can reach the interface with the magnet. The characteristic spin-transport lengths in these vertical spin-transport devices are often limited to nanoscale by scattering from magnetic impurities or by strong spin-orbit coupling. 

The limitation to nanoscale  lengths can be afforded in the vertical devices comprising thin-film stacks. However, the spatial-resolution scale of common electrical or optical probes that can be used to directly measure non-local spin-transport  between an injector and a detector in lateral structures is orders of magnitude larger. Correspondingly, also the fabrication of lateral spin-transistors and more complex spin-logic devices \cite{Datta1990,Koo2009,Wunderlich2010} relies on systems with spin-conserving length-scales approaching or, ideally, exceeding microns. The long-range nature of spin-transport, in combination with the desired high speed of these devices, inherently brings not only the requirement on the long spin life-time but also implies that electrons carrying the spin information were highly mobile.  

In semiconductors like GaAs, spin-polarized photo-carriers can be generated by optical spin orientation using circularly polarized light \cite{agranovich1984}. Scattering from magnetic impurities is then not relevant in these non-magnetic systems. Moreover, near the bottom of the conduction band of bulk GaAs, electron wavefunctions have a dominant $s$-orbital character which implies weak spin-orbit coupling.  The key factor limiting the spin life-time of photo-electrons is then the electron-hole recombination occurring at $\sim 100$~ps time-scales. This decay mechanism can be suppressed by introducing equilibrium carriers via impurity doping \cite{crooker2005,Dzhioev2002}. When electrons from the equilibrium electron sea take part in the recombination process, the net optically-induced  spin orientation can persist over $\sim 10-100$~ns after the pump pulse. The long spin life-time is achieved, however, at the expense of a low electron mobility ($\sim 10^3$~cm$^2$V$^{-1}$s$^{-1}$) in these bulk GaAs samples at dopings tuned close to the metal-insulator transition.

Approaching the problem from the opposite side, very high spin mobilities ($\sim 10^5-10^6$~cm$^2$V$^{-1}$s$^{-1}$) were reported \cite{wu2010,korn2010} in modulation doped two-dimensional electron gases (2DEGs) at GaAs/AlGaAs interfaces in which donors are introduced into the AlGaAs barrier and the electron sheet density of the GaAs two-dimensional channel  can be further controlled by gating \cite{gerlovin2007}. However, strong electric fields confining the 2DEG enhance the spin-orbit coupling which results in short spin life-times $\sim 100$~ps \cite{wu2010,korn2010}. 

Attempts to suppress spin relaxation in two-dimensional systems included growth on [110] or [111]-oriented GaAs substrates, instead of the common [001] orientation, resulting in a moderate spin life-time $\sim 1$~ns but low mobility $\sim 10^3$~cm$^2$V$^{-1}$s$^{-1}$ \cite{wang2013, hu2011}. An alternative approach is based on fine-tuning competing spin-orbit fields  in the confined two-dimensional systems to the so-called spin-helix state, or on quenching these fields in one-dimensional channels \cite{Wunderlich2009,Wunderlich2010,walser2012}. Moderate spin life-time $\sim 1$~ns and mobility $\sim 10^4-10^5$~cm$^2$V$^{-1}$s$^{-1}$ have been achieved in these studies \cite{walser2012}.

In our undoped GaAs/AlGaAs epilayer, electrons  have intrinsically high mobilty ($\sim 10^5$~cm$^2$V$^{-1}$s$^{-1}$ at 10~K)  since no intentional impurity doping is introduced into the system. The spin decay channel due to electron-hole recombination is suppressed by a  built-in electric field which drives photo-generated holes  towards the GaAs surface while the photo-electrons are driven in the opposite direction towards the AlGaAs barrier. The built-in field is not strong enough to confine the electrons into a narrow two-dimensional layer and to significantly enhance their spin-orbit coupling. As a result we observe spin life-times on the scale of 10's of ns. Our approach is reminiscent of the electron-hole separation reported in undoped, more complex structures comprising ZnSe/BeTe type-II quantum wells where, however, only a moderate spin life-time $\sim 1$~ns and mobility $\sim 10^4$~cm$^2$V$^{-1}$s$^{-1}$ were reported \cite{mino2011}.

Below we present our time and spatially-resolved pump-and-probe experiments in which we detect spins at distances reaching more than ten microns from the injection point at times as short as nanoseconds after the pump pulse. The corresponding long spin life-time and high mobility are independently confirmed by local time-resolved optical experiments and by measurements of the photo-electron conductivity in a Hall-bar dc-transport device and from terahertz (THz) transient conductivity spectra. 

\subsection{Optical time-resolved and spatially-resolved spin transport measurements}
Our pump and probe experiment, schematically illustrated in Fig.~1a, is performed on an undoped GaAs/AlGaAs structure grown by molecular beam epitaxy on a [001]-oriented semi-insulating GaAs substrate. The epilayers are deposited on the substrate starting from a 100~nm thick GaAs buffer, followed by a 100~nm Al$_{0.3}$Ga$_{0.7}$As barrier and a 800~nm thick top GaAs layer.  A circularly polarized pump beam and a time-delayed linearly polarized probe beam are focused at normal incidence on the sample surface, kept at temperature of 10~K, using a near infrared objective to form light spots of a diameter of $\sim2$~$\mu$m. The relative position of the pump and probe spots is varied by a piezo-tilting mirror. Absorption of the circularly-polarized pump light generates photo-electrons with out-of-plane spin-polarization.  An in-plane magnetic field, $B=500$~mT, is applied to induce Hanle precession of the photo-electron spins which are probed by time and spatially-resolved polar Kerr effect measurements. The time delay between two subsequent pump pulses is 12.5~ns. (For more details on the optical setup see Methods and Sec. S.2.1 in Supplementary information.)

\begin{figure}
\includegraphics[width=0.7\columnwidth]{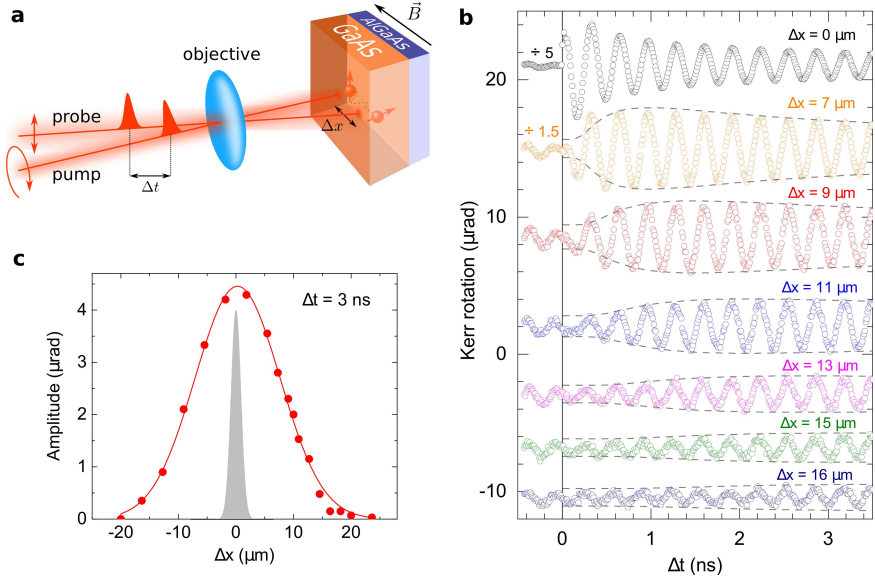}
\caption{\label{fig1} \textbf{Measurement of the spin life-time and spin diffusion coefficient.} \textbf{a}, A sketch of the spatially resolved pump-and-probe optical experiment. Circularly polarized pump and linearly polarized probe pulses are relatively time-delayed by $\Delta t$ and focused to 2~$\mu$m sized spots (full width at half maximum). The spots can be spatially separated by $\Delta x$ by tilting the incident pump beam. The magnetic field $B$ is applied in-plane along the pump-probe separation direction. The sample was kept at temperature 10~K. \textbf{b}, Kerr rotation dynamics measured by the spatially resolved pump and probe technique for several $\Delta x$. The $\Delta x$-dependent increase of the signal in time is a signature of diffusive spin-transport and is fitted by the non-oscillatory envelope from Eq.~\ref{eq4} (gray dashed curves). The data were vertically shifted and upper two curves were divided by indicated factors for clarity. \textbf{c}, Dependence of the signal amplitude  at $\Delta t =3$~ns with respect to $\Delta x$ (points: data, curve: fit using Eq.~\ref{eq4}). The spatial profile of the pump spot is depicted by the grey area.}
\end{figure}

The dynamics of the Kerr signal at varying spatial separation of the pump and probe laser spots, $\Delta x$, is shown in Fig.~\ref{fig1}b. The spin signal is detected at distances an order of magnitude larger than the laser spot size. This implies that the apparent decay of the signal with time at $\Delta x=0$ is only partly due to spin relaxation and that a significant contribution comes from the out-diffusion of the photo-injected spins from the area of the laser spot. Indeed, with increasing $\Delta x$, the character of the dynamic Kerr signal changes. It acquires a nearly time-independent amplitude between $\Delta t=1$ and 3~ns for $\Delta x=11$~$\mu$m and for larger $\Delta x$ it even increases with time. Consistently, we also observe a relative increase of the dynamic signal with increasing $\Delta x$ at negative times which originates from the previous pump-pulse. The spatial Gaussian evolution of amplitudes of the Kerr signal is compared to the pump beam profile in Fig.~\ref{fig1}c. Without a detailed analysis, the measured data illustrate that the spin signal  is transferred in our GaAs/AlGaAs structure over tens of microns in time intervals as short as nanoseconds. 

A more quantitative information on the spin life-time $\tau_s$ and diffusion coefficient $D_s$ of the photo-electrons, and on the corresponding spin diffusion length $l_s$, can be obtained by fitting our data to a solution of the two-dimensional spin diffusion equation derived for infinitesimally small injection and detection points \cite{crank1975,wu2010},
\begin{equation}
S(\Delta t,\Delta x) = \frac{N_s}{4\pi D_s \Delta t} \exp\left(-\frac{(\Delta x)^2}{4 D_s \Delta t}\right) \exp\left(-\frac{\Delta t}{\tau_s}\right) \cos(\omega \Delta t + \phi_0).
\label{eq4}
\end{equation}
Here $N_s$ is a normalization prefactor proportional to the total number of generated spins, $\omega=g\mu_B B/\hbar$ is the precession frequency, $g$ is the Land\'e g-factor, $\mu_B$ is the Bohr magneton, and $\phi_0$ is a phase factor related to the initial excitation conditions. For clarity, we show in Fig.~\ref{fig1}b only the envelope functions without the cosine factor. From the fits we obtain  $D_s = (130\pm30)$~cm$^2$s$^{-1}$ and $\tau_s = (20\pm 10)$~ns, and the corresponding spin diffusion length, $l_s = \sqrt{D_s \tau_s}=(16\pm4)$~$\mu$m. From the Einstein relation, $\mu_s=eD_s/k_B T$, and the temperature $T=10$~K of the experiment we can estimate the mobility of our photo-electron spins to $\mu_s\approx1.5\times10^{5}$~cm$^2$V$^{-1}$s$^{-1}$. (For a more detailed discussion of the $D_s$ vs. $\mu_s$ relation and the effect of finite size of the pump spot profile see Sec. S.2.4 in Supplementary information.)

\subsection{Optical measurements of the spin life-time}

\begin{figure}
\includegraphics[width=1.0\columnwidth]{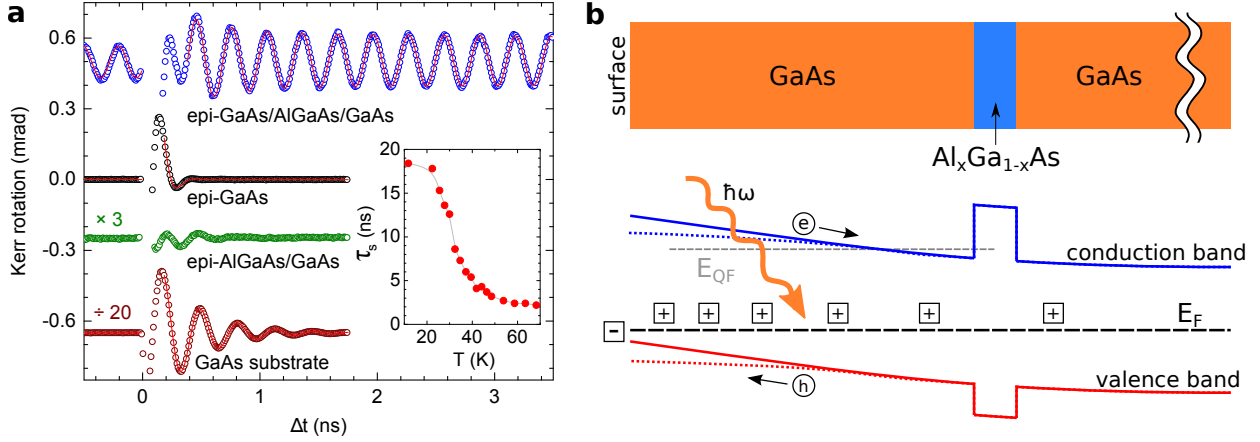}
\caption{\label{fig2} \textbf{Spin life-time $\tau_s$ and electron-hole separation.} \textbf{a}, Kerr rotation signals measured by the local pump and probe technique on the GaAs/AlGaAs/GaAs sample and three reference samples. The data were vertically shifted and bottom two curves were multiplied or divided by indicated factors for clarity. Inset: Temperature dependence of the measured spin life-time. \textbf{b}, Schematics of the GaAs/Al$_x$Ga$_{1-x}$As/GaAs layer structure  (sample surface is on the left side) and of the corresponding band-edge diagram. Data presented here were measured on samples with $x = 0.3$ and layer thicknesses 800~nm (top GaAs), 100~nm (Al$_x$Ga$_{1-x}$As) and 100~nm (bottom GaAs spacer) on 500~$\mu$m thick GaAs substrate. The band-edge diagram shows the band bending before the illumination (solid curves) due to the negatively charged surface states (square with the negative sign) and ionized unintentional impurity states in the bulk (squares with the positive signs); $E_{\text{F}}$ is the Fermi level. After the illumination by the pump light ($\hbar\omega$) the electron-hole pairs are created (e, h letters in circles) and split in the built-in electric field, resulting in the migration of photo-holes towards the surface and photo-electrons towards the barrier, and in the suppression of the band bending (dotted curves). The formation of the transient electron population near the interface is depicted by the quasi-Fermi level $E_{\text{QF}}$.}
\end{figure}

To confirm the large spin life-time inferred from the above non-local experiment we employ a local measurement with overlapping pump and probe light spots. The contribution from the out-diffusion of spins from the detection area is suppressed in this experiment by increasing the size of the spots to 25~$\mu$m. The typical time-dependent Kerr signal is shown in Fig.~\ref{fig2}a (blue points). Apart from additional signal components with a sub-ns spin life-time (commented further below and discussed in detail in Sec.~S.2.2 in Supplementary information), the spin life-time of the long-lived spin-system is obtained by fitting the Kerr dynamics to the expression \cite{sprinzl2010},
\begin{equation}
S(\Delta t) = S_0 \cos(\omega \Delta t + \phi_0)\exp\left(-\frac{\Delta t}{\tau_s}\right),
\label{eq1}
\end{equation}
where  $S_0$ is the initial amplitude of the signal for $\Delta t=0$. From the fit we get $\tau_s=(21\pm2)$~ns in agreement with the spatially-resolved measurement.

To highlight the large spin life-time observed in the GaAs/AlGaAs layer we present in Fig.~\ref{fig2}a also a measurement on a reference sample containing just the 900~nm thick GaAs epilayer on the GaAs substrate, and a measurement on the GaAs substrate alone. As expected,  spin life-times in these bulk undoped GaAs samples are limited to $\sim 100$~ps time-scales by the electron-hole recombination ($\tau_s\approx 70$~ps in the sample with the GaAs epilayer and $\tau_s\approx 350$~ps in the GaAs substrate).

The large $\tau_s$ observed in the GaAs/AlGaAs layer is exceptional for an undoped semiconductor. Its origin is illustrated in Fig.~\ref{fig2}b. In dark, GaAs surface states are negatively charged by electrons removed from unintentional impurity states in the bulk of the semiconductor \cite{sze2007,yablonovitch1989}. This generates a built-in electric field whose extent depends on the density of the ionized impurities and in epitaxial GaAs typically exceeds a micron. After illumination, photo-holes are driven by the built-in field towards the GaAs surface and fill the negatively charged surface states. Photo-electrons, on the other hand, partially fill the bulk impurity states in the top GaAs layer and partly accumulate at the interface of this layer with the AlGaAs barrier (for more details see Sec.~S.1 in Supplementary information). The resulting transient electron doping can persist over time-scales exceeding by orders of magnitude the repetition time of the pump pulses in  our optical measurements. We confirmed this by time-resolved THz conductivity measurements, discussed in the following section. 

As in previous experiments on GaAs structures with equilibrium excess electrons provided by intentional donor doping, our transient electron population  at the top GaAs/AlGaAs interface  suppresses the effect of electron-hole recombination on the spin life-time \cite{Dzhioev2002, gerlovin2007}. Simultaneously, due to the absence of intentional dopands in our structure, the electron mobility  is high  as we independently confirm below in the dc and THz conductivity measurements. Furthermore, consistently with the discussed mechanism, the electron-hole pairs generated outside the transient electron doping region (located near the interface) contribute to the total Kerr signal with a component whose spin-life time is limited by the electron-hole recombination to sub-ns time scale (see the rapidly damped part of the Kerr signal up to $\Delta t=1$~ns in Fig.~\ref{fig2}a, blue points).

The temperature dependence of the large $\tau_s$, shown in the inset of Fig~\ref{fig2}a, is consistent with the expected Dyakonov-Perel spin-dephasing mechanism dominating at higher temperatures and with hyperfine coupling to nuclear spins at low temperatures \cite{gerlovin2007}.  We also note that $\tau_s$ measurements as a function of the magnetic field strength confirm a uniform g-factor ($|g|=0.45\pm0.01$) in our structure which rules out a spin relaxation mechanism induced by the g-factor inhomogeneity. 

The picture of the electron-hole separation and of the accumulation  of electrons with large $\tau_s$ near the top GaAs/AlGaAs interface is further  supported by measurements on a structure without the top GaAs layer. As shown in Fig~\ref{fig2}a, the spin life-time in this sample (epi-AlGaAs/GaAs) is again on the $\sim 100$~ps time-scale. Indeed, the photo-holes generated in GaAs underneath the AlGaAs barrier cannot drift to the surface states and, consequently, the electron-hole separation mechanism is not effective at the interface between the AlGaAs barrier and the GaAs buffer layer.

\subsection{Dc-transport and THz-spectroscopy measurements of the electron mobility}
In dark, the undoped GaAs/AlGaAs layer is not conductive. Upon illumination, we detect transient dc-electrical conduction through a macroscopic Hall-bar device (a $14$~$\mu$m-long and $7$~$\mu$m-wide transport channel) by using both pulse laser-excitation with the 12.5~ns time separation between pulses or a continuous-wave Ti:Sapphire laser (see Sec.~S.3.2 in Supplementary information for data with cw-laser illumination). A long life-time of the electron system is independently inferred from the time-resolved  THz spectroscopy experiment with 200~$\mu$s delays between pump pulses (see Methods for details). Both the dc and THz experiments, which we now discuss in more detail, reveal a high mobility of the transient electron system, consistent with the large diffusion  constant observed above in the spatially-resolved spin transport measurements.

\begin{figure}
\includegraphics[width=0.8\columnwidth]{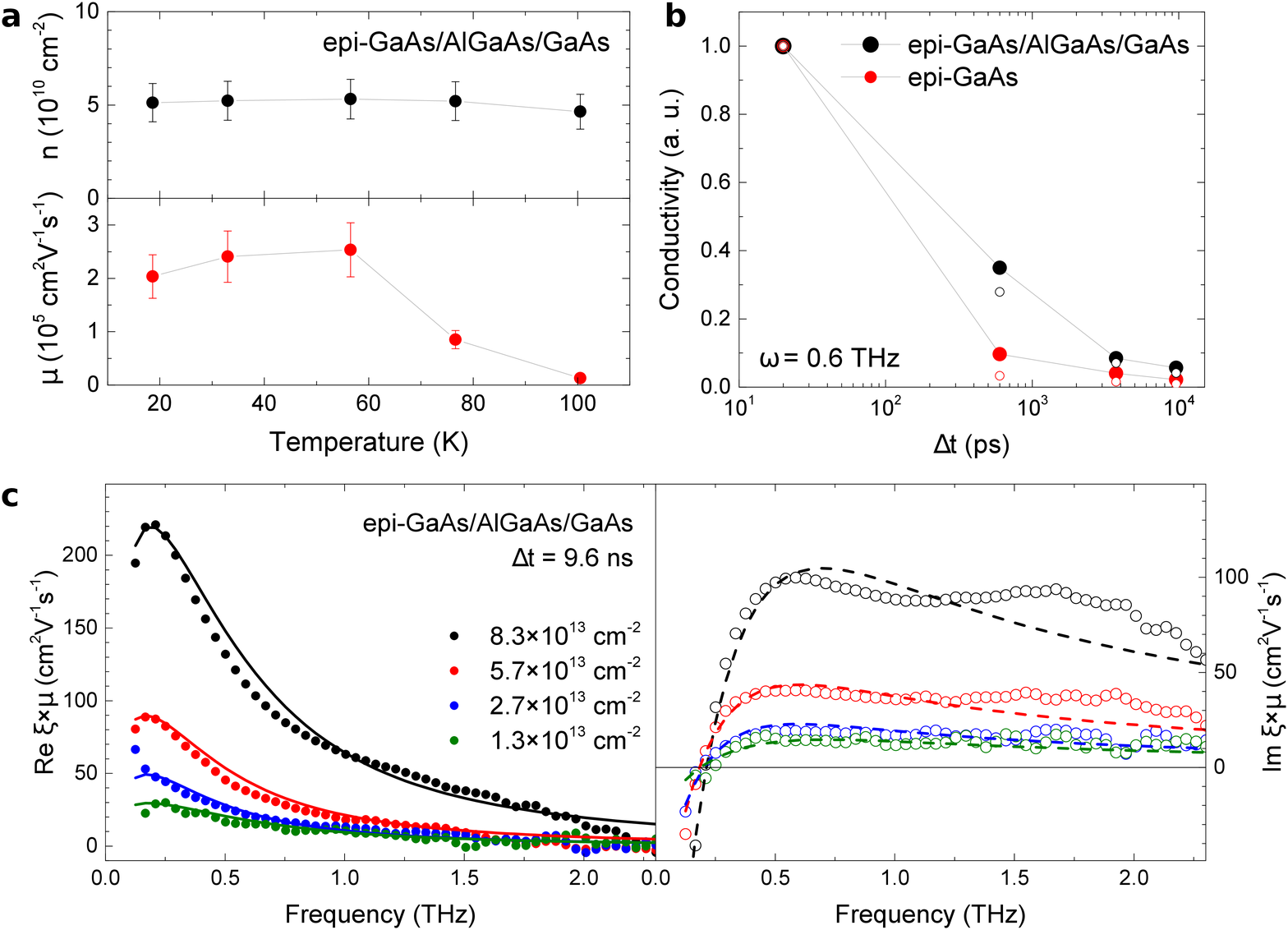}
\caption{\label{fig3} \textbf{Dc and THz transport measurements.} \textbf{a}, Dc Hall measurements of the transient carrier mobility and sheet carrier concentration with pulse laser illumination of the Hall cross at wavelength 815~nm and fluence generating a photocarrier density of $4\times10^{13}$~cm$^{-2}$. \textbf{b}, Time dependence of the representative value of the photoconductivity spectra (imaginary part at $\omega=0.6$~THz).  Data correspond to photoexcitation at fluence generating a carrier density of $2.6\times10^{13}$~cm$^{-2}$ and wavelength 800~nm (closed symbols) and at fluence generating a carrier density of $5.7\times10^{13}$~cm$^{-2}$ and wavelength 400~nm (open symbols), and to the GaAs/AlGaAs/GaAs sample (black symbols) and the reference sample containing the GaAs epilayer  on the GaAs substrate (red symbols). The photoconductivity at $\Delta t=20$~ps is normalized to unity. \textbf{c}, THz spectra of normalized complex photoconductivity as a function of pump photon fluence (i.e. of the density of photocarriers) at 400~nm and time delay 9.6~ns. Lines show fits by the two Drude terms from Eq.~\ref{eq3}.}
\end{figure}

The transient dc photoconductivity in our GaAs/AlGaAs layer is characterized at low temperatures by the electron mobility $\mu_n\approx 2\times 10^5$~cm$^2$V$^{-1}$s$^{-1}$ (data shown in Fig.~\ref{fig3}a bottom panel) and density $n\sim5\times 10^{10}$~cm$^{-2}$ (Fig.~\ref{fig3}a upper panel). (For more details see Sec.~S.3 in Supplementary information.) The electron mobility shows the characteristic bulk temperature dependence \cite{stillman1976} which confirms that the electric field confining the transient electron system near the GaAs/AlGaAs interface is relatively weak, as expected from the observed long spin life-time.  

The transient THz photoconductivity spectra $\Delta \sigma(\omega)$ were calculated from the transient wave form spectrum $\Delta E(\omega)$, measured at 18~K under various excitation fluences and wavelengths, using the expression \cite{Kuzel14},
\begin{equation}
\frac{\Delta \sigma(\omega)}{en_0}=\frac{1+n}{z_0}\frac{1}{e\phi}\frac{\Delta E(\omega)}{E(\omega)},
\label{eq2}
\end{equation}
where $n$ is the THz refractive index of GaAs, $z_0$ is the vacuum wave impedance, $\phi$ is the pump photon fluence absorbed in the sample, and $E(\omega)$ is the Fourier transform of the THz wave form transmitted through the unexcited sample (reference wave form). Here the conductivity is normalized by the initial charge density $en_0$, i.e., we obtain directly the conductivity per single unit charge. Note that this quantity is expressed in  units of the carrier mobility and it can be understood as a yield-mobility product $\xi(t)\mu(\omega)$, where $\mu(\omega)$ is the THz mobility spectrum, $\xi(\Delta t = 0) = 1$ is the unity quantum yield of the free carrier photo-excitation and $\xi(\Delta t)$ describes the time decay of the carrier density \cite{Kuzel14}. 

In Fig.~\ref{fig3}b we plot the time decay of the imaginary part of $\Delta \sigma$ at $\omega=0.6$~THz for a moderate photon pump fluence in the GaAs/AlGaAs/GaAs sample  and in the reference sample containing the GaAs epilayer on the GaAs substrate. Here, we first observe a significantly slower decay of the signal in the sample containing the barrier, compared to the reference sample, which indicates that the recombination process is dramatically changed when the AlGaAs barrier is present. This supports the picture of the in-build electric field affecting the electron-hole recombination channel by separating them. Measured data for photo-excitations at 800~nm and 400~nm, with the respective light penetration depths of $\sim 750$~nm and $\sim 20$~nm, also show similar time dependence of the THz conductivity which is a signature of the homogeneity of the electric field and of the similar efficiency of this mechanism for both excitation wavelengths.


Further evidence is provided by analyzing the fluence dependence of the THz spectra. Fig.~\ref{fig3}c shows an example of the spectra for  $\Delta t=9.6$~ns. We observe that the amplitude of the conductivity does not decrease with increasing pump fluence which would be the characteristics of a dominant  electron-hole recombination mechanism \cite{tanaka1983,kaiser1988}. (See Supplementary information Sec.~S.4.2 for the corresponding measurement in the reference bulk GaAs sample where electron-hole recombination dominates.) However, considering the life-time of electron-hole pairs in sub-ns time-scale seen in Fig.~\ref{fig3}b, we conclude that although the electron-hole recombination is obviously affected by the built-in electric field, the electron-hole separation of this part of the photo-carriers is not efficient enough to fully suppress their recombination. Consistently, this sub-system contributes with its sub-ns life-time to the Kerr signal in Fig.~\ref{fig2}a (blue points, data for $\Delta t<1$~ns).

Moreover, the imaginary part of the THz conductivity becomes negative at low frequencies and, correspondingly, the real part of the conductivity exhibits a maximum at a finite frequency, as required by Kramers-Kronig relations. This is a signature of another sub-system contributing to the net signal with a high-mobility term. The observation suggests that the highly mobile long-lived electron system is present before the arrival of each pump pulse (separated from the preceding pulse by 200~$\mu$s) and, therefore, it exhibits a fully suppressed electron-hole recombination. After being temporarily disturbed by the pump, this long-lived transient electron system leaves a trace in $\Delta\sigma(\omega)$ in the form of a negative Drude term in measured spectra. The presence of this steady-state highly mobile electron layer is fully consistent with the observation of the transient effective electron doping near the upper GaAs/AlGaAs interface in the time-resolved magneto-optical experiments.

To quantify the mobility of the long-lived electron system we fitted the THz spectra by a sum of positive and a negative  Drude terms,
\begin{equation}
\frac{\Delta \sigma(t, \omega)}{en_0}=\xi(t)\frac{e}{m^{\ast}}\frac{\tau}{1-i\omega \tau} - \xi_L\frac{e}{m^{\ast}}\frac{\tau_L}{1-i\omega \tau_L},
\label{eq3}
\end{equation}
where $\tau$ and $\tau_L$, and $\xi(t)$ and $\xi_L$ are the carrier scattering times and quantum yields of the respective positive and negative Drude terms. From the fits we obtain for the long-lived electron system $\tau_L= 3.3\pm0.7$~ps which corresponds to a carrier mobility $\mu_L=(1\pm0.2)\times10^5$~cm$^2$V$^{-1}$s$^{-1}$. This value is consistent, within a factor of 2, with the mobilities obtained from the dc Hall measurement and from the spatially-resolved optical spin transport measurement.

\subsection{Discussion}
We have demonstrated an exceptional  combination of a long spin life-time and a high mobility of an electron system in a semiconductor structure. Photo-generated electrons with the spin life-time reaching 20~ns and mobility $2\times10^5$~cm$^2$V$^{-1}$s$^{-1}$ have been demonstrated to allow for a transfer of the spin signal over distances of tens of microns  in times as short as nanoseconds. The simplicity of our bare GaAs/AlGaAs structure, compared to earlier attempts focusing on more elaborate doping, gating, heterostructure, or micro-patterning designs,  opens a new perspective for the research of spin transport in semiconductors. Since undoped GaAs/AlGaAs interfaces have been commonly embedded in a variety of previously studied structures, our results may  also prompt revisiting earlier experiments. As we highlight below, the superior spin-transport characteristics of the undoped  GaAs/AlGaAs interface is a robust and generic phenomenon.

\begin{figure}
\includegraphics[width=0.5\columnwidth]{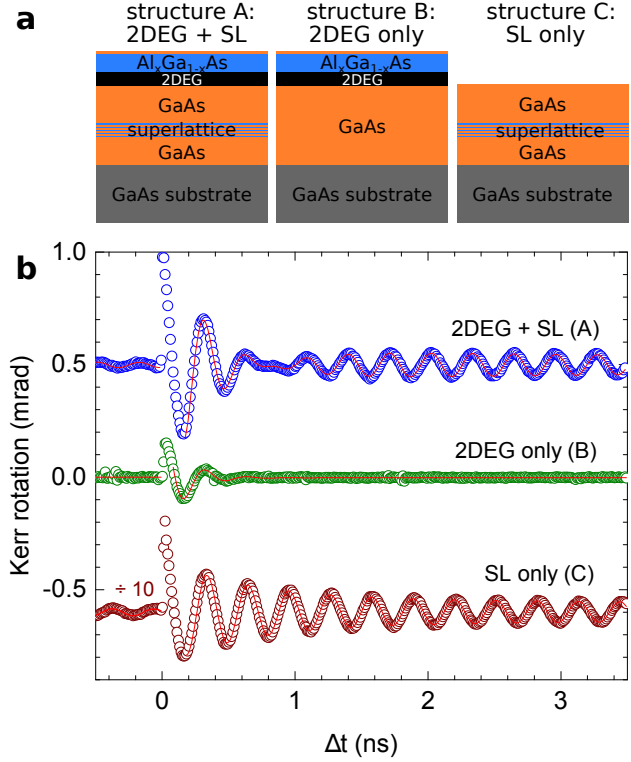}
\caption{\label{fig4} \textbf{Example of the robustness of the long spin life-time signal in complex heterostructures.} \textbf{a}, Schematics of the studied GaAs/AlGaAs heterostructures with a 2DEG formed in a triangular quantum well depicted by the black layer. \textbf{b}, Corresponding time-resolved Kerr rotation signals measured with the local pump and probe setup arrangement at same conditions as data presented in Fig.~\ref{fig2}a. The data were vertically shifted and divided by the indicated factor for clarity.}
\end{figure}

We have confirmed that the long spin life-time and high mobility are observed over a broad range of Al concentrations in the barrier and a broad range of barrier thicknesses (for more details see Sec.~S.2.3 in Supplementary information). Remarkably, the long spin life-time signal originating from photo-carriers at the GaAs/AlGaAs heterointerface prevails even in complex structures comprising multiple-period superlattices which are commonly introduced to separate doped GaAs quantum wells from the GaAs substrate. To demonstrate this, we compare in Fig.~\ref{fig4} optical spin life-time measurements in a series of GaAs/AlGaAs heterostructures. 

The structure (A) represents a common triangular quantum well system with a 2DEG residing in a GaAs layer grown underneath a doped AlGaAs barrier and separated from the GaAs substrate by an GaAs/AlAs superlattice. The time-resolved optical  signal measured in this structure is more complex than in the case of the bare GaAs/AlGaAs interface. Still, the long spin life-time component with $\tau_s\sim10$~ns is clearly present in the data. On the other hand, the structure (B) comprising an identical 2DEG layer but no undoped GaAs/AlAs superlattice shows only a short spin life-time signal with $\tau_s\sim100$~ps. Inversely, the structure (C) contains only the undoped superlattice and here the long spin life-time signal is again recovered, confirming the key role of the undoped GaAs/AlGaAs interface.

\section{Methods}
\subsection{Time and spatially resolved pump and probe experiment}
This experimental method was employed in order to measure $\tau_s$ and $D_s$ of the long-lived excess electron system. The sample was kept in an optical cryostat at temperature 10~K and subjected to in-plane magnetic field of 500~mT, provided by an electromagnet. A mode-locked Ti:Sapphire laser (Mai Tai, Spectra Physics) with repetition rate 80~MHz (i.e. time delay between two subsequent pulses of 12.5~ns) was tuned to the wavelength $\lambda=815$~nm. The laser beam was splitted to pump and probe beams with intensity ratio 10:1, setting the pump beam fluence to $\approx8.5$~$\mu$Jcm$^{-2}$, i.e. light power of 5~mW. This fluence corresponds to the bulk carrier density $\approx4.8\times10^{17}$~cm$^{-3}$ photo-generated in the sample, or to sheet carrier density $\approx4\times10^{13}$~cm$^{-2}$. Pump and probe beams were time delayed with respect to each other using a delay line, and then circularly and linearly polarized, respectively. The setup arrangement without the spatial resolution was used to determine $\tau_s$. Here, the incoming beams were focused using a lens into the cryostat close to the normal incidence on the sample surface, forming $\sim$25 $\mu$m-sized overlapped spots. 

In the arrangement with the spatial resolution (primarily employed to measure $D_s$), the lens was replaced by a near-infrared microscopic objective with magnification $20\times$ and numerical aperture 0.4 to create laser spots with full width at half maximum $<2$~$\mu$m. The pump beam focus point was movable in 2D using a piezo-tilting mirror holder. In order to suppress the reflected pump light in the detection part in this setup arrangement, the pump and probe beams were filtered using disjunctive spectral filters, resulting in central wavelengths 824 and 814~nm, respectively (more details can be found in Sec.~S.2.1 in Supplementary information). The reflected probe light was detected via an optical bridge in both arrangements.

\subsection{Electrical Hall experiment and THz spectroscopy}
These approaches were used in order to determine the carrier mobility of the long-lived system and to prove the spatial separation of electron-hole pairs. In the arrangement for the electrical Hall measurements, the sample was kept in an optical cryostat at temperature 18~K and the out-of-plane magnetic field of 500~mT was applied. The central detection area of the Hall cross nanopatterned on the sample surface was illuminated with the same pump laser beam and the same fluence as described in the preceding method. The electrical operation of the nominally undoped sample was possible due to the residual local optical illumination of whole sample surface (see Sec.~S.3 Supplementary information). The Hall cross was current-biased to 10~$\mu$A and the longitudinal and transversal voltages were sensed by dc voltmeters. 

The transient time resolved THz conductivity spectra were measured on unpatterned samples using a setup based on Ti:Sapphire laser amplifier with central wavelength 800~nm and 5~kHz repetition rate (time separation of two pulses is 200~$\mu$s). One part of the laser beam was used for the THz pulse generation and its phase-sensitive detection by means of the optical rectification and electro-optic sampling in 1-mm-thick (110)-oriented ZnTe crystals, respectively. Another part of the laser beam was used for optical pumping of samples either at 800 or 400 nm, see Ref.~\onlinecite{Fekete09} for more details. The pump beam was defocused to generate photocarriers homogeneously across the sample attached to a 3 mm aperture. The pump fluence was attenuated using neutral density filters to achieve the photo-generated carrier density in the range $\approx\times10^{12} - \times10^{14}$~cm$^{-2}$. An optical chopper was used in the pump beam path; this scheme ensures that we experimentally detect the photo-induced change $\Delta E(\omega)$ of the THz wave form transmitted through the sample. The experiments were carried out at 18 K for both pump wavelengths.

\section{Acknowledgements}
We acknowledge support from the European Research Council (ERC) Advanced Grant No. 268066,
from European Metrology Research Programme within the Joint Research Project EXL04 (SpinCal), 
from the Ministry of Education of the Czech Republic Grant No. LM2011026,
from the Czech Science Foundation Grant No. 13-12386S,
from the Grant Agency of the Czech Republic Grant No. 14-37427G,
from the Charles University Grants No. 1360313 and No. SVV-2015-260216.

\section{Author contributions}
L.N., P.N., K.O., H.N., P.K., J.W. and T.Ju. designed the experiment and interpreted the data. L.N. and T.Ju. wrote the manuscript. L.N. performed the spatially and time-resolved magneto-optical, the electrical Hall experiments and the nano-lithography, V.S. and P.K. carried out the THz experiments, V.N. grew the samples, T.Ja. and F.T. carried out the remote control of the setup. 



\providecommand{\noopsort}[1]{}\providecommand{\singleletter}[1]{#1}%

\end{document}